\documentclass[twocolumn]{aastex631}
\usepackage{amsmath}

\newcommand{\be}{\begin{equation}}
\newcommand{\ee}{\end{equation}}

\shorttitle{Sensitivity of the 
Hubble Constant Determination to Cepheid Calibration}
\shortauthors{M\"ortsell et al.}

\begin{document}

\title{Sensitivity of the 
Hubble Constant Determination to Cepheid Calibration}

\correspondingauthor{Edvard M\"ortsell}
\email{edvard@fysik.su.se}

\author{Edvard M\"ortsell}
\affiliation{Oskar Klein Centre, Department of Physics, Stockholm University\\Albanova University Center\\ 106 91 Stockholm, Sweden}

\author{Ariel Goobar}
\affiliation{Oskar Klein Centre, Department of Physics, Stockholm University\\Albanova University Center\\ 106 91 Stockholm, Sweden}

\author{Joel Johansson}
\affiliation{Oskar Klein Centre, Department of Physics, Stockholm University\\Albanova University Center\\ 106 91 Stockholm, Sweden}

\author{Suhail Dhawan}
\affiliation{Institute of Astronomy\\
University of Cambridge
Madingley Road \\ Cambridge CB3 0HA\\
United Kingdom}



\begin{abstract}
Motivated by the large observed diversity in the properties of
extra-galactic extinction by dust, we re-analyse the Cepheid
calibration used to infer the Hubble constant,
$H_0$, from Type Ia supernovae, using Cepheid data in 19 
Type Ia supernova host galaxies from \citet{Riess:2016jrr} and
anchor data from \citet{Riess:2016jrr, Riess_2019, Riess_2021}.
Unlike the SH0ES team, we do not
enforce a fixed universal color-luminosity relation to correct the 
Cepheid magnitudes. Instead, we focus on a data driven method, where
the optical colors and near infrared magnitudes of the Cepheids are used to derive individual color-luminosity relations for each Type Ia supernova host and anchor galaxy. We present two different analyses, one based on Wesenheit magnitudes resulting in $H_0=73.2\pm 1.3$ km/s/Mpc, a $4.2\,\sigma$
tension with the value inferred from the cosmic microwave background.
In the second approach, we calibrate an individual extinction law for each galaxy with non-informative priors using color excesses, yielding $H_0=73.9\pm 1.8$ km/s/Mpc, in $3.4\,\sigma$ tension with the Planck value. 

Although the two methods yield similar results, in the latter approach the 
Hubble constant inferred from the individual Cepheid absolute distance calibrator galaxies range from  {$H_0=68.1\pm 3.5$} km/s/Mpc to {$H_0=76.7\pm 2.0$} km/s/Mpc. Taking the correlated nature of $H_0$ inferred from individual anchors into account and allowing for individual extinction laws, the Milky Way anchor is in $2.1\,\sigma - 3.1\,\sigma$ tension with the NGC 4258 and the Large Magellanic Cloud anchors, depending on prior assumptions regarding the color-luminosity relations and the method used for quantifying the tension.
\end{abstract}



\section{Introduction}
As is well known, there is a tension between the value of the Hubble constant as inferred from small and large distance measurements, most significantly between the values inferred from Type Ia supernova (SNIa) distances to redshifts $z\sim 0.1$ calibrated by Cepheid observations, as measured by the SH0ES team, and the distance to the cosmic microwave background (CMB) decoupling surface at $z\sim 1090$, as measured by the Planck satellite. The former yields $H_0=73.2\pm 1.3$ (in units of km/s/Mpc used from now on) \citep{Riess_2021} and the latter $H_0=67.4\pm 0.5$ \citep{Planck2020}; a $4.1\,\sigma$ tension\footnote{During the referee process, the SH0ES team have presented an updated Hubble constant value of $H_0=73.04\pm 1.04$ ($4.9\,\sigma$ tension) using an, yet not publicly available, expanded SNIa and Cepheid data set \citep{Riess:2021jrx}.}

The tension between other measurements is not as significant: Calibrating the absolute SNIa magnitude using the tip of the red giant branch (TRGB) observations gives $H_0=69.6\pm 1.6$ \citep{Freedman_2019}, right between the Cepheid calibrated SNIa and the CMB inferred values. 

Another local estimate of the expansion rate of the Universe is derived from the amplitude of the gravitational wave signal GW170817, the merger of a binary neutron-star system located to the galaxy NGC 4993 at $z=0.010$  through the electromagnetic counterpart, AT2017gfo,  yielding a Hubble constant with relatively large uncertainties of $H_0 = 70.0^{+12.0}_{-8.0}$ \citep{Abbott:2017xzu}.

\begin{figure}[ht]
    \centering
	\includegraphics[width=1\linewidth]{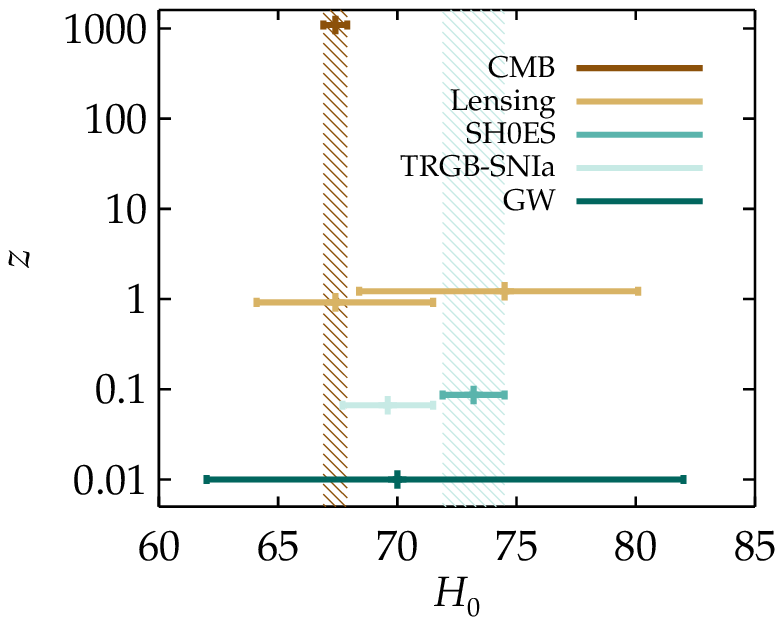}
	\caption{The Hubble constant as inferred from distances measured to the CMB, strongly lensed quasars, SNIa calibrated with Cepheids (SH0ES) and the TRGB, and the gravitational wave signal GW170817. The redshift corresponds to the mean redshift of the distances employed in the method, slightly shifted where needed to avoid overlap. The major tension is between the values inferred from Cepheid calibrated SNIa by the SH0ES team and CMB observations. 
	\label{fig:Tension}}
\end{figure}

A different route involves gravitational lensing. Time delays in the TDCOSMO sample of seven lensed quasars yield $H_0=74.5^{+5.6}_{-6.1}$ \citep{Birrer_2020}. Constraining the galaxy lens mass profiles, using kinematics observations of an independent set of gravitational lenses in the Sloan Lens ACS sample (SLACS), lowers the value to $H_0=67.4^{+4.1}_{-3.2}$, assuming that the TDCOSMO and SLACS galaxies are drawn from the same parent population. These results are illustrated in Figure~\ref{fig:Tension}, from which is evident that only the Cepheid calibrated SNIa 
distance scale from SH0ES is in definite tension with the CMB inferred distance. 

The value inferred from the CMB depends on the entire expansion history of the Universe, whereas the SNIa measurement only depends on the local expansion rate it sets out to measure. On the other hand, the inference from SNe Ia depends on a combination of a larger number of astrophysical probes. Therefore, attempts to modify the CMB inferred $H_0$ usually rely on modifications of the cosmological model, whereas the SNIa value is usually studied with emphasis on possible systematic effects concerning the local distance measurements.  

At least in principle, the CMB value can be increased by various departures from the concordance cosmological constant and cold dark matter, $\Lambda$CDM model, see e.g. \citet{Mortsell_2018,Knox_2020}. Options include decreasing the physical size of the sound horizon used to measure the distance to $z=1090$. This can be accomplished by adding sources of energy present before CMB photon decoupling, e.g., new thermal relativistic species or early dark energy, or by reducing the sound speed. However, such modifications are severely constrained when taking the full CMB power spectrum into account. 
Attempts to shift the CMB value also involve changing the expansion history at redshifts $z<1090$, with modest success since the expansion rate is tightly constrained by SNIa and baryonic acoustic oscillation (BAO) observations.

Given that the proposals mentioned above require substantial modifications of the current concordance cosmological model, and still fail in relieving the full tension, we investigate the Cepheid-SNIa value and its uncertainties, see also \citet{Follin_2018,efstathiou2020lockdown}. In particular, we concentrate on dust extinction, affecting all astronomical observations in the optical and near infrared (NIR) regime. 
We focus on a a very specific assumption made by the SH0ES team throughout their series of publications, namely that there is a fixed {\em universal} reddening law in all galaxies. 

\section{Revisiting dust extinction corrections}\label{sec:revisit}
In spite of the critical importance for precision cosmology, the current understanding of light attenuation in the interstellar medium (ISM) of galaxies remains very limited. In comparison, the Milky Way (MW) ISM has been studied in great detail, including the properties of dust grains responsible for dimming of light\citep[see][for a review]{2003ARA&A..41..241D}. In particular, several MW reddening laws have been devised, among these \citet{1989ApJ...345..245C} (CCM), \citet{1994ApJ...422..158O} and \citet{1999PASP..111...63F} (F99).
They have in common the use of a single parameter, the total to selective extinction coefficient, $R_{\rm V}^{\rm BV}$, as a proxy for the grain composition and size distribution, where the attenuation in the optical ${\rm V}$-band relates to the color excess $E({\rm B}-{\rm V})\equiv A_{\rm B}-A_{\rm V}$ as $A_{\rm V} = R_{\rm V}^{\rm BV}\,E({\rm B}-{\rm V})$, here referred to as the "CCM-relationship". Low values of $R_{\rm V}^{\rm BV}$ indicate a steep wavelength dependence, while large $R_{\rm V}^{\rm BV}$ correspond to gray extinction.

While an average $\langle R_{\rm V}^{\rm BV} \rangle=3.1 $ for the MW is found in most studies, significant variations are found in individual lines of sight in the galaxy, ranging from $R_{\rm V}^{\rm BV} \approx 2$ in some diffuse sight lines, to $R_{\rm V}^{\rm BV}\approx 6$ in dense molecular clouds \citep{1999PASP..111...63F}. 
The diversity in the MW has been confirmed by \citet{2016MNRAS.456.2692N}, who find significantly lower values of $R_{\rm V}^{\rm BV}$ in the Galactic bulge. Moving the scope outside the MW, a study by \citet{2003ApJ...594..279G} of the extinction in the Magellanic Clouds found that a small number of Large Magellanic Cloud (LMC) extinction curves are consistent with the CCM relationship, but the majority of the LMC and all the SMC curves are not. 
\citet{2015MNRAS.450.3597F} report a gray extinction law for NGC 4258 in the line of sight of the Cepheids, $R_{\rm V}^{\rm BV} \sim 4.9$, although they caution that this could be the result of unresolved systematics. 
For more distant galaxies, observed SNIa colors highlight the observed diversity in extinction properties, ranging from $R_{\rm V}^{\rm BV} \sim 1$ to values consistent with the MW average  \citep[see][and references therein]{2006AJ....131.1639K,2008A&A...487...19N, 2014ApJ...784L..12G,2014ApJ...788L..21A,2015MNRAS.453.3300A,2018ApJ...869...56B}. For the SNe~Ia in the Hubble flow, color corrections are based on the SALT2 lightcurve fitter \citep{2007A&A...466...11G}, which again differ from the CCM parameterization, but are most consistent with values of $R_{\rm V}^{\rm BV}\sim 2.5$ \citep[see e.g.,][and references therein]{2021arXiv210316978B}, although dust extinction 
differences between host galaxy environments has been suggested as an explanation for a systematic "mass step" in the derived distances \citep{2021ApJ...909...26B,2021arXiv210506236J}. 

\begin{figure}[ht]
    \centering
	\includegraphics[width=1\linewidth]{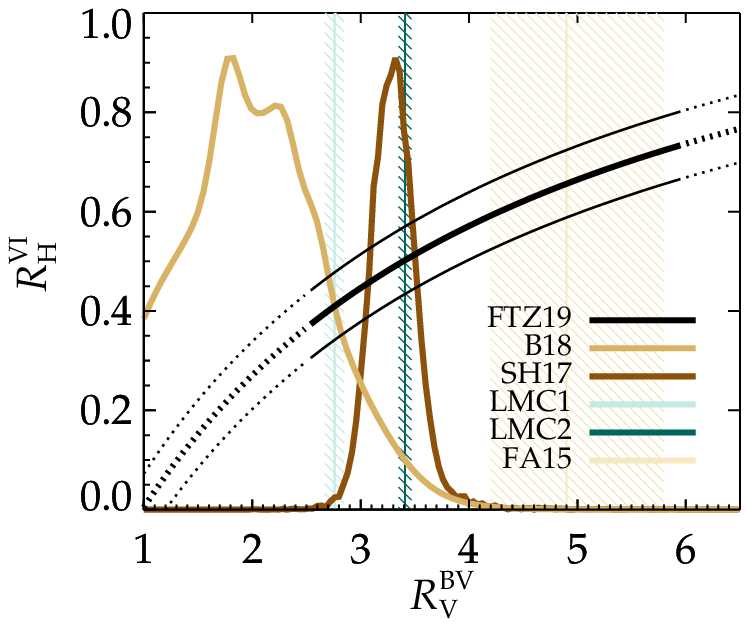}
	\caption{$R_{\rm H}^{\rm VI}$ vs $R_{\rm V}^{\rm BV}$ relation and  1$\sigma$ scatter from the extinction law derived by  \citet{2019ApJ...886..108F} (FTZ19), extrapolated to $R_{\rm V}^{\rm BV}>6$ and $R_{\rm V}^{\rm BV}<2.5$. Such low values have been 
inferred from SNIa colors \citep[][B18]{2018ApJ...869...56B}. Also shown are derived  $R_{\rm V}^{\rm BV}$ values from a MW stellar sample in \citet{2017ApJ...838...36S} (SH17), two samples in the LMC \citep[][LMC1 and LMC2]{2003ApJ...594..279G} and in the water mega-maser anchor NGC 4258 \citep[][FA15]{2015MNRAS.450.3597F}.}
	\label{fig:RVvsRHfun}
\end{figure}

To minimize the impact from extinction correction uncertainties, the SH0ES team use flux measurements in the NIR H-band, centered at 1.6 $\mu m$, where extinction by dust, gauged using the observed color ${\rm V}-{\rm I}$, is significantly smaller. Adopting the CCM-like relationship from F99 and the extinction correction $A_{\rm H} = R_{\rm H}^{\rm VI}\,E({\rm V}-{\rm I})$, the value corresponding to the MW average is $R_{\rm H}^{\rm VI}\sim 0.4$.  However, there is no theoretical, nor any empirical studies of extinction suggesting that a {\em universal} value of $R_{\rm H}^{\rm VI}$ can be assumed. On the contrary, a recent study by  \citet{2019ApJ...886..108F} finds considerable variations between lines of sight for the extinction curves in the NIR in the MW. Based on a parameterization fitting extinction laws from ultraviolet to NIR for 72 well-measured stars, a very wide range in $R_{\rm H}^{\rm VI}$ can be inferred, as shown in Figure~\ref{fig:RVvsRHfun}.
Hence, assuming a narrow range in $R_{\rm H}^{\rm VI}$ for the anchor and Cepheid hosts in not warranted by current observations.
In this paper, we investigate to what degree relaxing this assumption affects the inferred value of $H_0$ and its corresponding uncertainties. 

\section{Difference in methodology}
The local distance ladder uses the difference between Cepheid magnitudes in SNIa hosts and anchor galaxies. Therefore, it is relatively insensitive to changes in the {\em global} properties of the extinction law.
In \citet{Riess:2016jrr}, a global change of the parameter $R_{\rm W}$ parameterizing a Cepheid color-luminosity (C-L) correction with respect to the observed color $R_{\rm W}\,({\rm V}-{\rm I})$ of $R_{\rm W}=0.39\rightarrow 0.35$ changed the Hubble constant by $\delta H_0\sim 0.5$. When multiplied with the observed color ${\rm V}-{\rm I}$, the parameter $R_{\rm W}$ is argued to also partly correct for an intrinsic Cepheid C-L relation.

In \citet{Follin_2018}, a slightly different C-L correction with respect to an estimated color excess of $R_{\rm E}\,\hat{E}\,({\rm V}-{\rm I})$ was employed, where $\hat{E}\,({\rm V}-{\rm I})\equiv ({\rm V}-{\rm I})-\langle{\rm V}-{\rm I}\rangle_0$ with $\langle{\rm V}-{\rm I}\rangle_0$ an estimate of the mean intrinsic Cepheid color. Here, $R_{\rm E}$ is interpreted as the dust total to selective extinction ratio, $R_{\rm H}^{\rm VI}$. Imposing a prior of $R_{\rm E}=0.39\pm 0.1$, a value of $H_0 = 73.3\pm 1.7$ was derived when allowing $R_{\rm E}$ to vary between galaxies, in good agreement with $H_0=73.2\pm 1.3$ from \citet{Riess_2021}. 
At face value, this result seems to suggest that the method used for calibrating the C-L relation and/or the possibility of varying this calibration between galaxies have a small impact on the inferred Hubble constant. 

Since dust extinction is uncertain at NIR wavelengths, we investigate the effect of allowing for $R_{\rm W}$ and $R_{\rm E}$ to be fitted by the Cepheid data. Since dust properties are also known to vary between galaxies, when calibrating with respect to color excesses, we allow $R_{\rm E}$ to do the same. Given the lack of solid independent constraints on $R_{\rm E}$, we employ less restrictive priors than \citet{Follin_2018}. Any systematic difference in dust properties or the intrinsic C-L between SNIa hosts and anchor galaxies will shift the inferred value of $H_0$. 

\section{Method and Data}
The apparent magnitude of a source 
at redshift $z$ with absolute magnitude $M$ is given by
\be
m=5\log D(z) + M + 25,
\ee
where $D(z)$ is the luminosity distance in units of Mpc. By combining observed magnitudes of Cepheids in SNIa host and anchor galaxies, $m_{\rm Ceph}^{\rm host}$ and  $m_{\rm Ceph}^{\rm anch}$ with SNIa magnitudes in host galaxies and in the Hubble flow,  $m_{\rm SN}^{\rm host}$ and  $m_{\rm SN}^{\rm flow}$ , we can derive
\be
5\log H_0 = 5\log r(z)-5\log D^{\rm anch}+\Delta m_{\rm SN}-\Delta m_{\rm Ceph},
\ee
where $r(z)\equiv H_0 D(z)$ can be approximated by $r(z)\approx cz$ in the close Hubble flow and we have defined 
\begin{align}
\Delta m_{\rm SN}& \equiv m_{\rm SN}^{\rm host}-m_{\rm SN}^{\rm flow},\\
\Delta m_{\rm Ceph}& \equiv m_{\rm Ceph}^{\rm host}-m_{\rm Ceph}^{\rm anch}.
\end{align}

Apart from getting the SNIa redshifts and anchor distances right, we thus need to make sure there are no systematic offsets in the Cepheid and SNIa magnitudes between host, anchor and cosmic flow galaxies. Ignoring the weak cosmology dependence of $r(z)$ \citep{Dhawan_2020}, the inferred value of $H_0$ will decrease (increase) if we:
\begin{enumerate}
\item Increase (decrease) the independent anchor distances, $D^{\rm anch}$.
\item Decrease (increase) $\Delta m_{\rm SN}$.
\item Increase (decrease) $\Delta m_{\rm Ceph}$.
\end{enumerate}
Here, we focus on option 3. With regards to option 2, $\Delta m_{\rm SN}$ will increase if SNIa in Cepheid host galaxies are systematically made brighter than in the Hubble flow, e.g., if there is additional dust extinction not accounted for in the host galaxies, or if the effect that SNIa in high mass hosts are systematically brighter than in low mass galaxies, such as Cepheid hosts, have been underestimated \citep[see][and references therein]{2020A&A...644A.176R}.

In terms of option 3, if there is additional dust extinction not accounted for in the anchor galaxies, or if we have over-corrected for dust extinction in the host galaxies, the inferred value of $H_0$ will decrease, and vice versa. The fractional shift in the Hubble constant is
\be
\frac{\delta H_0}{H_0}= \frac{\delta r(z)}{r(z)}-\frac{\delta D^{\rm anch}}{D^{\rm anch}}
	+\frac{\ln 10}{5}\left[\delta(\Delta m_{\rm SN}) -\delta(\Delta m_{\rm Ceph})\right],
\ee
and a lower limit to the precision in $H_0$ is set by the precision of the anchor distance measurements. Shifting $\delta (\Delta m_{\rm Ceph})=\pm 0.1$ will shift the Hubble constant by $\delta H_0\mp 4.6\,\%$. 

\subsection{Cepheid calibration}
We use the Hubble Space Telescope (HST) flux in the NIR filter (${\rm H}={\rm F160W}$) band, color calibrated using optical (${\rm V}={\rm F555W}$ and ${\rm I}={\rm F814W}$) data, to derive Wesenheit magnitudes
\be\label{eq:whorig} 
m_{\rm H}^{\rm W} \equiv m_{\rm H} - R_{\rm W}\,({\rm V} - {\rm I}) =  m_{\rm H} - R_{\rm W}\,E({\rm V} - {\rm I}) - R_{\rm W}\,({\rm V} - {\rm I})_0,
\ee 
where the color excess $E\,({\rm V} - {\rm I})\equiv A_{\rm V}-A_{\rm I}=({\rm V} - {\rm I})-({\rm V} - {\rm I})_0$, with $({\rm V} - {\rm I})_0$ the intrinsic Cepheid color.
In the last step, we see that $m_{\rm H}^{\rm W}$ is corrected both for dust extinction, identifying the first $R_{\rm W}$ with the total to selective extinction ratio $R_{\rm H}^{\rm VI}\equiv A_{\rm H}/(A_{\rm V}-A_{\rm I})$, and for a possible intrinsic C-L relation, identifying the second $R_{\rm W}$ with $\beta_{\rm H}^{\rm VI}$, as parameterized in e.g. \citet{1982ApJ...253..575M} and \citet{2017ApJ...842...42M}. The term $\beta_{\rm H}^{\rm VI}\,({\rm V} - {\rm I})_0$ corresponds to the intrinsic magnitude-color relation at a fixed Cepheid period, whereas the correlation between the intrinsic color with period is included in the period-luminosity (P-L) calibration parameterized by $b_{\rm W}$ in equation~\ref{eq:mHW} below. 
As an alternative approach, also employed in \citet{Follin_2018}, we calibrate the C-L relation using
\be\label{eq:whevi} 
m_{\rm H}^{\rm W} \equiv m_{\rm H} - R_{\rm E}\,\hat{E}\,({\rm V} - {\rm I}).
\ee
Here, $\hat{E}\,({\rm V} - {\rm I})$ represents a proxy for the color excess obtained by subtracting an estimate of the mean intrinsic colors, $\langle{\rm V} - {\rm I}\rangle_0$ from the observed colors\footnote{Note that if $\langle{\rm V} - {\rm I}\rangle_0$ is assumed to depend on the Cepheid period, the fitted $b_{\rm W}$ parameterizing the P-L relation will shift accordingly.},
\be
\hat{E}\,({\rm V} - {\rm I})\equiv ({\rm V} - {\rm I})-\langle{\rm V} - {\rm I}\rangle_0.
\ee
The estimated color excess $\hat{E}\,({\rm V} - {\rm I})$ also represents a combination of dust extinction and intrinsic color, since the mean intrinsic colors, $\langle{\rm V} - {\rm I}\rangle_0$ does not take into account individual variations in Cepheid temperature along the width of the Cepheid instability strip, see e.g. \citet{1991PASP..103..933M,2006ARA&A..44...93S,Pejcha_2012}. 
Using multi wavelength data, one can in principle attempt to distinguish the contribution from dust and intrinsic color variations, see e.g. \citet{Pejcha_2012,2017ApJ...842...42M} and calibrate them separately.
In the following, we will follow standard practice and assume that $R_{\rm W}$ and $R_{\rm E}$ effectively corrects both for dust and intrinsic color variations, noting that the former will dominate since temperature variations are subdominant given the narrow width of the instability strip. Taking an empirical approach, for the calibration with respect to the observed color we will fit for the value of $R_{\rm W}$ that minimizes the scatter in $m_{\rm H}^{\rm W}$. When calibrating with respect to color excesses, we will allow for individual galactic $R_{\rm E}$, representing a variation in the dust properties between galaxies.

We model the Wesenheit magnitude of the $j$th Cepheid in the $i$th SNIa host as
\be\label{eq:mHW}
m_{{\rm H},i,j}^{\rm W }=\mu_i+M_{\rm H}^{\rm W}+b_{\rm W}[{\rm P}]_{i,j}+Z_{\rm W} [{\rm M}/{\rm H}]_{i,j},
\ee
where $[{\rm M}/{\rm H}]_{i,j}$ is a measure of the metallicity of the Cepheid, $[{\rm P}]_{i,j}\equiv \log P_{i,j} -1$ where $P_{i,j}$ is the period measured in days, $M_{\rm H}^{\rm W}$ the absolute Cepheid magnitude normalized to a period of $P=10$ days and Solar metallicity and $\mu_i$ the distance modulus to the $i$th galaxy. In what follows, we will allow for separate P-L relations for short and long period Cepheids using 
\be
b_{\rm W}[{\rm P}]_{i,j}\rightarrow b^{\rm s}_{\rm W}[{\rm P}]^{\rm s}_{i,j}+ b^{\rm l}_{\rm W}[{\rm P}]^{\rm l}_{i,j},
\ee
where $[{\rm P}]^{\rm s}_{i,j}=0$ for Cepheids with periods $>10$ days and  $[{\rm P}]^{\rm l}_{i,j}=0$ for Cepheids with periods $<10$ days, see Section~\ref{sec:sysofeqns}.

Similarly for the $j$th Cepheid in the $k$th anchor galaxy, here MW, NGC 4258 and the LMC, 
\be 
m_{{\rm H},k,j}^{\rm W}=\mu_k+M_{\rm H}^{\rm W}+b_{\rm W}[{\rm P}]_{k,j}+Z_{\rm W} [{\rm M}/{\rm H}]_{k,j}.
\ee

\subsection{Milky Way Cepheids}
Trigonometric parallaxes potentially provide the most direct calibration of the Cepheid absolute magnitude, $M_{\rm H}^{\rm W}$.
We use data from \citet{Riess_2021}, with 75 MW Cepheids, out of which 68 have reliable GAIA parallaxes. 
For the $j$th Cepheid in the MW, 
\be 
m_{{\rm H},j}^{\rm W}=\mu_j+M_{\rm H}^{\rm W}+b_{\rm W}[{\rm P}]_{j}+Z_{\rm W} [{\rm M}/{\rm H}]_{j}.
\ee
where the distance modulii for each Cepheid is estimated using GAIA parallaxes, $\pi$, according to 
\be
\pi_j + zp=10^{-0.2(\mu_j-10)},
\ee
where $zp$ is a residual parallax calibration offset that we fit for together with $M_{\rm H}^{\rm W}, b_{\rm W}$ and $Z_{\rm W}$. In \citet{Riess_2021}, $M_{\rm H}^{\rm W}$ and $zp$ are fit for using only MW data setting $b_{\rm W} = -3.26$ and $Z_{\rm W}=-0.17$ (as fitted to all Cepheids), finding $zp=-14\pm 6\,\mu{\rm as}$. Since we want to fit for $zp$ simultaneously with all parameter, we write
\begin{align}
\mu_j&=10-\frac{5}{\ln 10}\left[\ln\pi+\ln\left(1+\frac{zp}{\pi}\right)\right]\nonumber \\& = 10-\frac{5}{\ln 10}\left[\ln\pi+\frac{zp}{\pi}+\mathcal O \left(\frac{zp}{\pi}\right)^2\right],
\end{align} 
effectively transforming $zp$ into a linear parameter, and
\begin{align}\label{eq:mpi}
m_{{\rm H},j}^{\rm W}-10+\frac{5}{\ln 10}\ln\pi& =m_{\rm H}^{\rm W}+b_{\rm W}[{\rm P}]_{j}\nonumber \\
&+Z_{\rm W} [{\rm M}/{\rm H}]_{j}-\frac{5}{\ln 10}\frac{zp}{\pi}.
\end{align} 
Higher order terms, $\mathcal O(zp/\pi)^2$, are small and corrected for in an iterative manner.

\subsection{Type Ia Supernovae}
The calibrated SNIa B-band peak magnitude in the $i$th host is modelled by
\be 
m_{{\rm B},i}=\mu_i+M_{\rm B}.
\label{eq:snmag} 
\ee
The SNIa peak apparent magnitudes need to be corrected for the width-luminosity and C-L relations. There are several lightcurve fitting algorithms for deriving the SNIa peak magnitude, lightcurve shape and color, the most widely used for cosmology being the SALT2 model \citep{2007A&A...466...11G}. The derived lightcurve widths and colors are used to correct the peak magnitude $m_{\rm B}$ in equation~\ref{eq:snmag}. The errors on the corrected peak magnitude include the fitting error and a 0.1 mag term from the SNIa model added in quadrature.

\subsection{Data}
For the extra-galactic (M31 and beyond) Cepheids, including Cepheids in the anchor galaxy NGC 4258, we use the data set from Table~4 in \citet{Riess:2016jrr}. This table is restricted to Cepheids passing a best-fit, global $2.7\,\sigma$ outlier rejection, the impact of which is claimed to be small in \cite{Riess:2016jrr}, but not possible to confirm independently by us.

For Cepheids in the LMC, we use data in Table~2 in \citet{Riess_2019}. Data for MW Cepheids, including GAIA parallax measurements are from Table~1 in \citet{Riess_2021}.

Double eclipsing binaries (DEBs) provide a means to measure distances by determining the physical sizes of the member stars via their radial velocities and light curves \citep{Paczynski:1996dj}. 
20 DEBs observed using long-baseline near-infrared interferometry give a distance to the LMC of $\mu_{\rm LMC}=18.477\pm 0.0263$ \citep{Pietrzy_ski_2019,Riess_2019}.
We use the updated distance to NGC 4258 of $\mu_{\rm N4258}=29.397\pm 0.032$ \citep{Reid_2019}, using observations of mega-masers in Keplerian motion around its central super massive black hole.

Type Ia SN B-band magnitudes are from Table~5 in \citet{Riess:2016jrr}, derived using version 2.4 of SALT II \citep{Betoule_2014}.

\subsection{Parameter Fitting}\label{sec:paramfit}
Given the observed Cepheid magnitudes $m_{\rm H}$, colors ${\rm V} - {\rm I}$, periods $[{\rm P}]$, metallicities $[{\rm M}/{\rm H}]$, together with the SNIa magnitudes $m_{\rm B}$, the anchor distances $\mu_k$ and the MW Cepheid parallaxes $\pi$, we can fit simultaneously for $R_{\rm W}$ or $R_{\rm E}$, $b_{\rm W}$, $Z_{\rm W}$, the host galaxy distances $\mu_i$, the anchor distances $\mu_k$, the GAIA parallax offset $zp$, the Cepheid absolute magnitude $M_{\rm H}^{\rm W}$ and the SNIa absolute magnitude $M_{\rm B}$. For linear parameters, the fit can be made analytically as described in Section~\ref{sec:sysofeqns}. Although $R_{\rm W}$ and $R_{\rm E}$ appear as linear parameters in equation~\ref{eq:whorig} and \ref{eq:whevi}, since the uncertainty in the observed Cepheid colors are non-negligible, the uncertainty in the derived Wesenheit magnitudes $m_{\rm H}^{\rm W}$ will depend on the values of $R_{\rm W}$ or $R_{\rm E}$, requiring a non-linear treatment of these parameters, see Section~\ref{sec:sysofeqns}. Effectively, a proper color uncertainty treatment will assign a larger Wesenheit magnitude uncertainty for larger values of $R_{\rm W}$ and $R_{\rm E}$, whereas not taking the color correction uncertainty into account will bias the result towards too low values of $R_{\rm W}$ and $R_{\rm E}$. Whether this will bias the inferred Hubble constant low or high depends on the relative bias between $R_{\rm E}$ in SNIa host and Cepheid distance calibrator (anchor) galaxies. For illustrative purposes, we will present an example of the magnitude of this bias in the Results sections.

Given $M_{\rm B}$, the Hubble constant is calculated as 
\be 
H_0=10^{M_{\rm B}/5+a_{\rm B}+5} 
\label{eq:h0} 
\ee
where $a_{\rm B}$ is the intercept of the SNIa magnitude-redshift relation
\begin{align} 
10^{a_{\rm B}+m_{\rm B}/5}&=cz \bigg\{1 + \frac{1}{2}\left[1-q_0\right] {z}\nonumber \\
&-\frac{1}{6}\left[1-q_0-3q_0^2+j_0 \right] z^2 + O(z^3)\bigg\}
\label{eq:ax} 
\end{align}
measured with $q_0=-0.55$ and $j_0=1$ to $a_{\rm B}=0.71273 \pm 0.00176$ \citep{Riess:2016jrr}.

\section{Results for the Wesenheit calibration \texorpdfstring{$R_{\rm W}\,({\rm V} - {\rm I})$}{xxx}}
In \citet{Riess_2021}, a value of $H_0=73.2\pm 1.3$ is derived combing anchor distances from the MW, NGC 4258 and LMC.

Using the same data with $R_{\rm W}=0.386$, using double P-L relations, doing a full count-rate non-linearity correction following \citet{2019wfc..rept....1R}, identifying $[{\rm O}/{\rm H}]=[{\rm Fe}/{\rm H}]$ using $Z_\odot=8.824$, and fitting for the residual GAIA offset $zp$ simultaneously with all other parameters, we obtain $H_0 = 73.1 \pm 1.3$.
Here, we have added a scatter in the P-L relation of $\sigma(m_{\rm H}^{\rm W})=0.0682$, to give a $\chi^2/{\rm dof}=1$. 
Despite slight differences in the analysis method, this value is in very good agreement with the value in \citet{Riess_2021}, and in $4.1\, \sigma$ tension with the Planck value of  $H_0=67.4\pm 0.5$ \citep{Planck2020}.

\subsection{Fitting for \texorpdfstring{$R_{\rm W}$}{xxx}}
The fact that dust extinction extrapolated to the H-band is very uncertain naturally opens up for the option of fitting for the value of $R_{\rm W}$. Assuming  a global value common for all galaxies gives $R_{\rm W}= 0.37 \pm 0.02$ with $H_0 = 73.2\pm 1.3$ ($4.2\,\sigma$ tension). Here, we have used a flat prior of $R_{\rm W}=[0,1]$, but given the small posterior uncertainty on $R_{\rm W}$, results are insensitive to the specific choice of this prior.

\subsection{Individual P-L relations}
\begin{figure}[ht]
    \centering
	\includegraphics[width=1\linewidth]{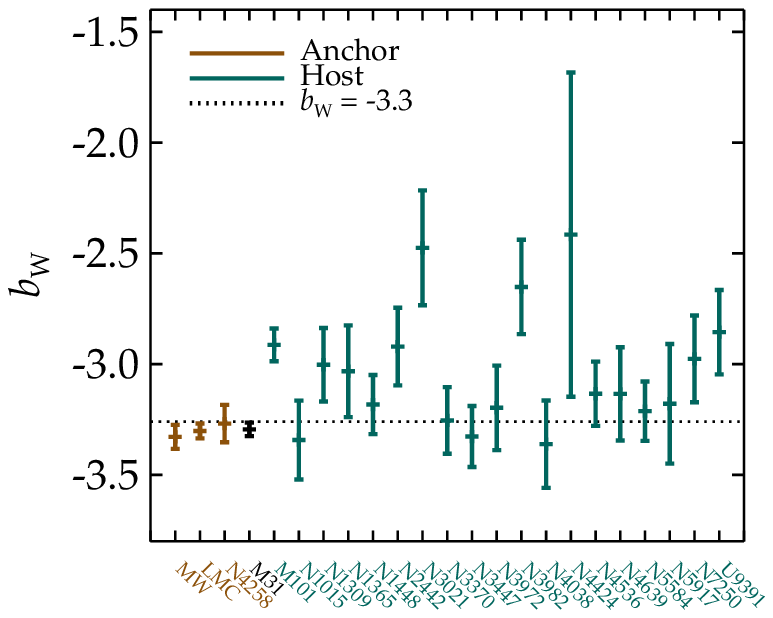}
	\caption{Fitting individual P-L relations $b_{\rm W}$ to Cepheid data for a fixed global value of $R_{\rm W}=0.386$,. Anchor galaxies are denoted in brown and SNIa host galaxies in petrol. The dotted line corresponds to $b_{\rm W}=-3.26$, indicating the value obtained assuming a global P-L relation.
	\label{fig:indbW}}
\end{figure}

So far, we have assumed that all Cepheids can be described by a global P-L relation, described by $b_{\rm W}^{\rm s}$ and $b_{\rm W}^{\rm l}$. However, since there is evidence that the P-L relation can vary between galaxies \citep{Tammann_2011,efstathiou2020lockdown}, in a similar spirit to our approach of allowing $R_{\rm W}$ to vary between galaxies, we investigate to what extent relaxing this assumption will affect the inferred Hubble constant. We will allow for individual galactic values of $b_{{\rm W},i}$ to be fitted for simultaneously with all other parameters, in this case restricting to the same P-L relation for short and long period Cepheids. For a fixed global value of $R_{\rm W}=0.386$, the resulting $b_{{\rm W},i}$ are shown in Figure~\ref{fig:indbW}. The fact that the fitted $b_{{\rm W},i}$ are systematically higher in SNIa host galaxies compared to hosts, the inferred Hubble constant is increased from $H_0 = 73.1\pm 1.3$ to $H_0 = 76.5\pm 1.9$. 
If we allow for {\em both} individual P-L and a globally fitted C-L relation, the Hubble constant is again $H_0 = 76.5\pm 1.9$, with Planck tension $4.7\,\sigma$.

\section{Results for color excess calibration 
\texorpdfstring{$R_{\rm E}\,\hat{E}({\rm V} - {\rm I})$}{xxx}}
We next compare with results derived when color calibrating the Cepheid sample with respect to the estimated color excess.  
We derive $\hat{E}\,({\rm V} - {\rm I})$ subtracting mean intrinsic colors as estimated in \citet{Tammann_2011}, including the quoted uncertainties and a $0.075$ dispersion in the mean intrinsic Cepheid color between galaxies (inferred from the difference between LMC and MW Cepheids). These colors are in good agreement with results in \citet{Pejcha_2012}. The intrinsic color uncertainties are included generating random Monte Carlo samples. The mean color excess range from $\hat{E}({\rm V} - {\rm I})=0.28$ in the LMC to $\hat{E}({\rm V} - {\rm I})=0.69$ in the MW, see also Figure~\ref{fig:PervsVI_R16}.
Calibrating using $R_{\rm E}\,\hat{E}({\rm V} - {\rm I})$, for a fixed value of $R_{\rm E}=0.386$, we obtain $H_0 = 73.0 \pm 1.3$, showing the insensitivity of calibration method for fixed values of $R_{\rm W}$ and $R_{\rm E}$.

\begin{figure}[ht]
    \centering
	\includegraphics[width=1\linewidth]{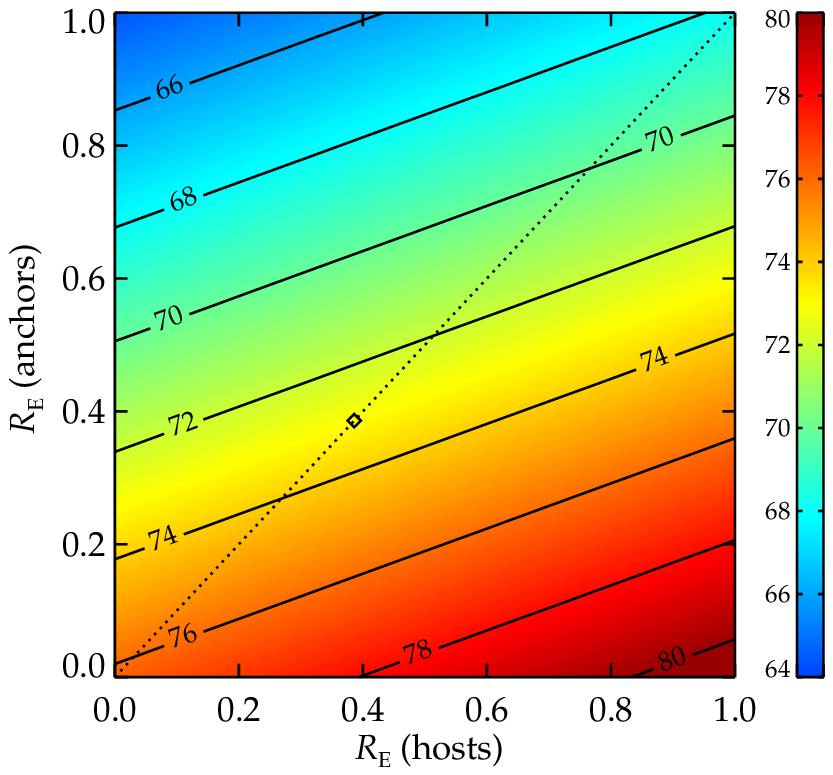}
	\caption{$H_0$ as a function of $R_{\rm E}$ in the SNIa hosts and the anchor galaxies when color calibrating Cepheids with respect to the estimated color excess $R_{\rm E}\,\hat{E}({\rm V} - {\rm I})$. The diamond indicates $R_{\rm E}=0.386$ as assumed in \citet{Riess:2016jrr, Riess_2019, Riess_2021}. 
	\label{fig:H0grid_intcol}}
\end{figure}

The inferred value of $H_0$ will shift if there is a systematic offset in $R_{\rm E}$ between anchor and SNIa host galaxies. 
We derive values for $H_0$ when varying the value of $R_{\rm E}$ in the anchor(s) and the SNIa hosts, see Figure~\ref{fig:H0grid_intcol}. For a common $R_{\rm E}$ (indicated by the dotted line), the inferred value of $H_0$ decreases when $R_{\rm E}$ is increased. Also, $H_0$ is decreased when $R_{\rm E}$ is larger in anchor than in host galaxies.
Given the result in Figure~\ref{fig:H0grid_intcol}, one could argue that a simple solution to the Hubble tension, would be a systematic shift of $R_{\rm E}$ between anchor and host galaxies, or an increased overall value of $R_{\rm E}$ bringing $H_0$ as inferred from SNIa down to the Planck value. This could be argued for, e.g., if dust properties in SNIa host galaxies have a systematically steeper extinction law than the anchor galaxies. In lack of solid independent evidence for such a systematic shift, or the value of $R_{\rm E}$ and possible variations of it, we will next include $R_{\rm E}$ as model parameters to be constrained by the available data.

\subsection{Fitting for \texorpdfstring{$R_{\rm E}$}{xxx}}
Fitting for a global value of $R_{\rm E}$ (again using a flat prior $R_{\rm E}=[0.1]$, common for all galaxies, gives $R_{\rm E}= 0.37 \pm 0.02$ with $H_0 = 73.2\pm 1.7$ with Planck tension $3.3\,\sigma$.
Allowing also for individual galactic P-L relations, we obtain $H_0 = 76.0\pm 2.2$ ($3.9\,\sigma$ tension).

\begin{figure}[ht]
    \centering
	\includegraphics[width=1\linewidth]{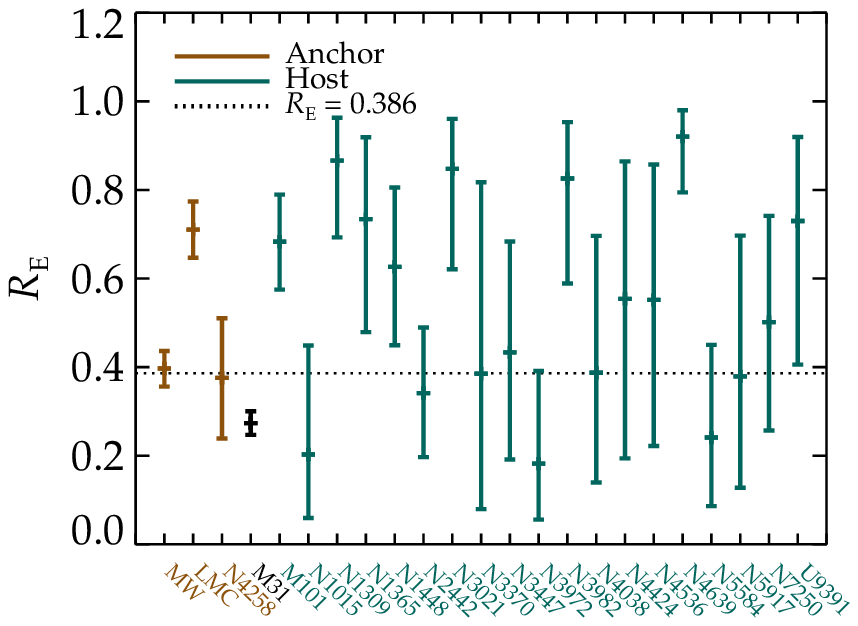}
	\caption{The result of fitting individual galactic values for $R_{\rm E}$ using the color excess Cepheid calibration, imposing flat prior constraints $R_{\rm E}=[0,1]$. Anchor galaxies are denoted in brown and SNIa host galaxies in petrol. The dotted line corresponds to $R_{\rm E} = 0.386$.
	\label{fig:rhmin00max10}}
\end{figure}

We next allow for $R_{\rm E}$ to vary between galaxies. 
Since $R_{\rm E}$ primarily represent corrections to dust, with large uncertainties in the NIR, we use wide priors of $R_{\rm E}=[0,1]$ as a default to investigate the full impact of using the data at hand to constrain $R_{\rm E}$. We obtain $H_0=73.9\pm 1.8$, in $3.4\,\sigma$ tension with the Planck value, see Figure~\ref{fig:rhmin00max10}. 
If not properly taking into account the dependence of the Wesenheit magnitude uncertainties on the $R_{\rm E}$, as discussed in Section~\ref{sec:paramfit}, one obtains 
$H_0=72.2\pm 1.7$ ($2.7\,\sigma$ tension), i.e., a bias of $\Delta H_0=1.7$.

Imposing a set of more restrictive flat priors $R_{\rm E}=[0.15,0.8]$, the inferred Hubble constant is $H_0=73.6\pm 1.8$. Our most restrictive set of priors is derived by assuming that the (mean) reddening of extra-galactic Cepheids is well-represented by the reddening distribution of MW stars. Using the sample of $R_{\rm V}^{\rm BV}$ from $\sim 15\,000$ MW stars in \citet{2017ApJ...838...36S} (see Figure~\ref{fig:RVvsRHfun}), converted to $R_{\rm H}^{\rm VI}$ using the results in \citet{2019ApJ...886..108F}, the derived distribution is well-approximated by a Gaussian with $R_{\rm E}=0.48\pm 0.1$. For this set of priors, $H_0=73.4\pm 1.7$. The corresponding sets of posterior $R_{\rm E}$ are shown in Figure~\ref{fig:rhmin015max08}.

Allowing also for individual galactic P-L relations, we obtain $H_0 = 75.2\pm 2.4$ ($3.2\,\sigma$ tension), assuming the wide priors on $R_{\rm E}=[0,1]$. 

At face value, these results suggest that the inferred Hubble constant is quite insensitive to the Cepheid color calibration scheme. In terms of the results for individual anchors however, when allowing for individually fitted $R_{\rm E}=[0,1]$, the Hubble constants inferred from each individual anchor distance show a substantial spread with the MW anchor distance being dominant in pushing the combined Hubble constant to the high value in tension with Planck measurements, see  Section~\ref{sec:indanch}. Since the different $H_0$ derived for each individual anchor distance are correlated, the significance of the tension can not be immediately read out from the $H_0$ and their corresponding uncertainties. 

Instead, we quantify the significance of differences between anchors in the following ways. First, using {\em only} MW Cepheids, we derive the Cepheid absolute magnitude, $M_{\rm H}^{\rm W}$. We next compare each of the independently measured  $\mu_{\rm LMC}$ and $\mu_{\rm N4258}$ to the values derived using the other two anchors. For example, using NGC 4258 and LMC as anchors but no MW Cepheids, we compare the derived value of $M_{\rm H}^{\rm W}$ to that previously obtained using only MW Cepheids. Also, using the MW Cepheids value for $M_{\rm H}^{\rm W}$ and the LMC distance as anchors, we compare the derived $\mu_{\rm N4258}$ to the independently derived value. In order to account for the "look elsewhere" effect, we make use of a Monte Carlo approach where we simulate three independent data points from the same mean value, and arbitrary variances. Using a large number of realisations, we compare each of the three points to the weighted mean of the other two points. We compute the fraction of realizations that has a certain maximum deviation for the three data points to quantify how a given value for the observed significance corresponds to a slightly lower significance of the results. For the case of $R_{\rm W}=0.386$, NGC 4258 shows the largest deviation of $2.0\,\sigma$, corresponding to a modest $1.6\,\sigma$ significance accounting for the "look elsewhere" effect. For individually fitted $R_{\rm E}=[0,1]$, the MW anchor has a deviation of $3.0\,\sigma$, corresponding to a $2.6\,\sigma$ significance.

A drawback of this method when quantifying the tension of the MW anchor is that using only the 67 MW Cepheids, the uncertainty of the inferred $M_{\rm H}^{\rm W}$ is fairly large (we typically obtain $M_{\rm H}^{\rm W}=-5.88\pm 0.04$ for the Wesenheit calibration and $M_{\rm H}^{\rm W}=-5.57\pm 0.05$ when color calibrating with respect to estimated intrinsic colors). As an alternative approach, using only the MW Cepheids as distance anchors, we infer the distances to NGC 4258 and the LMC and compare to their independently measured values. Despite the slight differences between the two approaches, they yield broadly consistent results for the MW anchor tension. Results are summarized in Figure~\ref{fig:anchtensionwide} and Table~\ref{tab:anchtension} for different priors on the color-luminosity relation. 
For the color excess calibration, the MW anchor is in $2.1\,\sigma - 3.1\,\sigma$ tension with the NGC 4258 and the LMC anchors, depending on the assumed prior distributions of $R_{\rm E}$. Note again that we are here using data from \citet{Riess:2016jrr, Riess_2019, Riess_2021}. In \citet{Riess:2021jrx}, an expanded data set is used to show good consistency between the individual anchor distances for a fixed value of $R_{\rm W}=0.386$, attributed to a refined Cepheid metallicity dependence decreasing the inferred correction term from $Z_{\rm W}=-0.13\pm 0.07$ \citep{Riess:2016jrr}
to $Z_{\rm W}=-0.22\pm 0.05$.

\section{Summary and Discussion}
\begin{figure}[!t]
    \centering
	\includegraphics[width=1\linewidth]{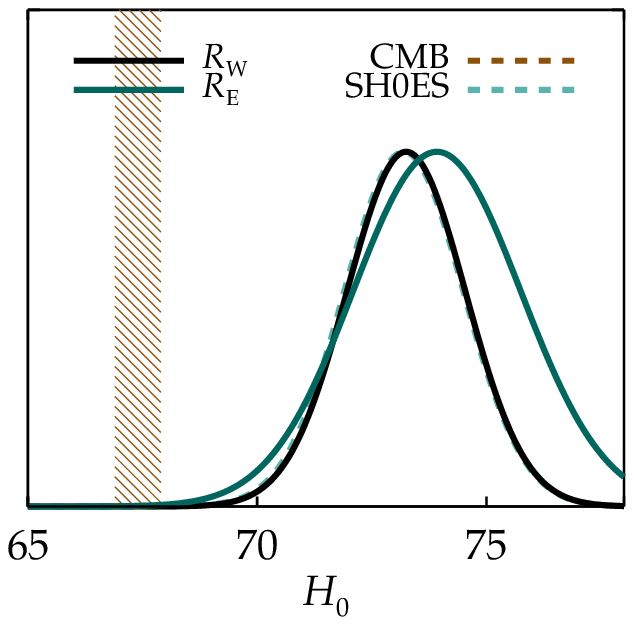}
	\caption{
	Results for $H_0$ for the two main analyses employed in this paper. The solid black line is for a fitted global value of $R_{\rm W}$ using Wesenheit magnitudes, overlapping with the dashed petrol line fitted using the Wesenheit calibration with $R_{\rm W}= 0.386$ as in \citet{Riess_2021}.
	The dark petrol line is for individual $R_{\rm E}$ using color excesses, assuming prior values $R_{\rm E}=[0,1]$. The dashed brown region indicates the $1\,\sigma$ region from Planck \citep{Planck2020}.
	\label{fig:Finalresult}}
\end{figure}

We have investigated the sensitivity of the Hubble constant inferred from SNIa distance measurements to the choice of Cepheid calibration method. Specifically, we have compared results when color calibrating the Cepheid magnitudes with respect to observed colors and estimated color excesses. Guided by the lack of independent evidence for dust extinction properties at NIR wavelengths, we have allowed for the color calibration parameters to be determined by the Cepheid data; either a global value or in the case of the color excess calibration, individual galactic values.

For the color excess calibration, we derive $H_0=73.9\pm 1.8$ in $3.4\,\sigma$ tension with the value inferred from Planck, using priors of $R_{\rm E}=[0,1]$. Using more restrictive priors on $R_{\rm E}$ only has small effects on the inferred Hubble constant.
Calibrating with respect to observed colors yields $H_0 = 73.2\pm 1.3$, see Figure~\ref{fig:Finalresult}. Allowing also for individual galactic P-L relations, the corresponding $H_0$ values are increased to $H_0 = 75.2\pm 2.4$ and $H_0 = 76.5\pm 1.9$, respectively. 

Results for different calibration choices are summarized in Table~\ref{tab:IC} in Section~\ref{sec:modelselection}. From the quality of the fits, there is no clear preference for any of the calibration methods with different information criteria showing preference for different amount of freedom in the model.  

\begin{figure}[!ht]
    \centering
	\includegraphics[width=1\linewidth]{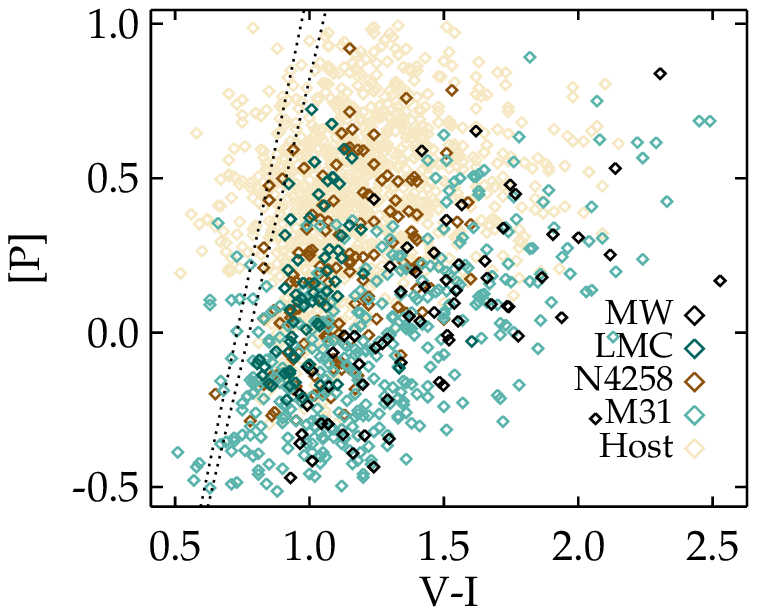}
	\caption{Observed Cepheid colors ${\rm V} - {\rm I}$ versus the periods $[P]=\log P -1$. The dotted lines correspond to the upper and lower limit of the mean intrinsic MW Cepheid color as estimated in \citet{Tammann_2011}.
	\label{fig:PervsVI_R16}}
\end{figure}

\begin{figure}[!ht]
    \centering
	\includegraphics[width=1\linewidth]{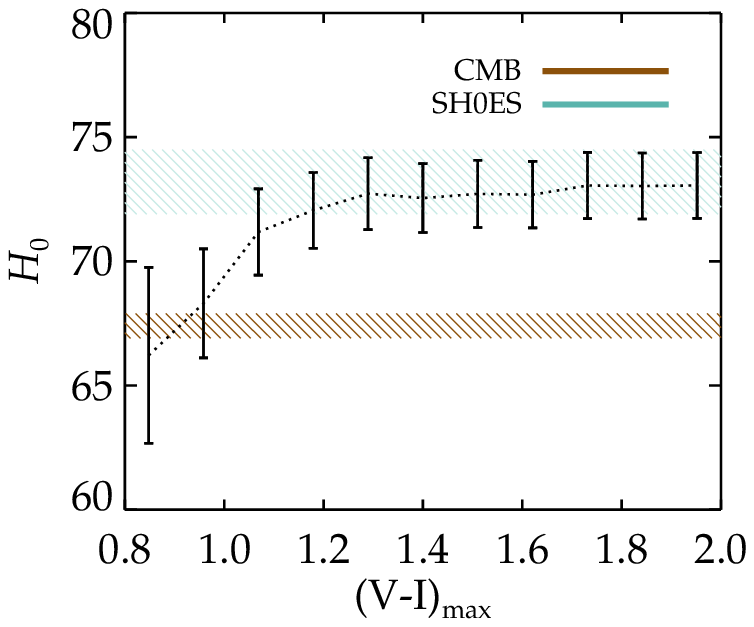}
	\caption{Fitted $H_0$ as a function of the cut in $({\rm V} - {\rm I})$ for a fixed value of $R_{\rm E}=0.386$ when calibrating using estimated color excesses.
	\label{fig:h0vsvicut_intcol}}
\end{figure}

Regardless of calibration method, since Cepheid colors and periods are correlated \citep{Tammann_2011}, so will the inferred P-L and C-L relations, i.e., the parameters $b_{\rm W}$ and $R_{\rm W}$ or $R_{\rm E}$. This may be of importance if there are color selection effects related to the fact that longer period Cepheids are brighter, see Figure~\ref{fig:PervsVI_R16}. We have tested the possible impact of such an effect by imposing cuts on the observed color ${\rm V} - {\rm I}$. In Figure~\ref{fig:h0vsvicut_intcol}, we show the fitted $H_0$ as a function of the cut in $({\rm V} - {\rm I})$ we apply, when calibrating using color excesses for a fixed $R_{\rm E}=0.386$. 
With $({\rm V} - {\rm I})_{\rm max}=[2, 1.5, 1.25, 1]$, $[98\,\%, 85\,\%, 65\,\%, 29\,\%]$ of the original Cepheids remain. As evident from Figure~\ref{fig:h0vsvicut_intcol}, only for $({\rm V} - {\rm I})_{\rm max}\approx 1$ does the cut significantly degrade the statistical uncertainty in $H_0$.
The trend of obtaining a lower $H_0$ when cutting out redder Cepheid is a common feature for all calibration methods and color-luminosity priors we have tested. 

Finally, we note that when estimating distances using the ${\rm I}$-band for which color corrections are larger, the sensitivity of the result to the choice of calibration method is larger. For example, with the Wesenheit calibration, assuming a fixed value of $R_{\rm W,I}^{\rm VI}=1.27$ (corresponding to $R_{\rm H}^{\rm VI}=0.386$), we obtain $H_0 = 70.4\pm 1.5$, in $1.9\,\sigma$ tension with the CMB value, whereas fitting for a global value of $R_{\rm W,I}^{\rm VI}$ yields $H_0 = 71.4\pm 1.3$ (a $2.8\,\sigma$ tension).
With the color excess calibration, for $R_{\rm E,I}^{\rm VI}=1.27$, we obtain $H_0 = 69.7\pm 1.5$, in $1.4\,\sigma$ tension with the CMB value, whereas allowing for individual galactic values yields $H_0 = 69.8\pm 1.8$, with $1.2\,\sigma$ Planck tension \footnote{Using flat priors $R_{\rm E,I}^{\rm VI}=[0.215,2.95]$ corresponding to $R_{\rm E}=[0,1]$}.


With the limited information at hand regarding dust extinction for Cepheids at NIR wavelengths, the color calibration of Cepheid magnitudes could potentially introduce large uncertainties in the local distance ladder. Allowing for a global $R_{\rm W}$ or individually fitted values of $R_{\rm E}$ in the anchor(s) and the SNIa hosts does not significantly change the inferred value of $H_0$, although the $H_0$ as derived from the individual anchors will shift. In the case of individually fitted values of $R_{\rm E}$, the inferred values range from $H_0 = 68.1\pm 3.5$ for the NGC 4258 anchor to  $H_0 = 76.7\pm 2.0$ for the Milky Way. Neither approach employed in this paper is in one-to-one correspondence with an underlying physical model and there is no clear evidence in the data for any of them.

\begin{acknowledgements}
We thank the anonymous referee for the insightful and thorough reviews of the manuscript, in particular for pointing out the importance of a proper treatment of color errors having substantial impact on the results when fitting for the values of $R_{\rm W}$ and $R_{\rm E}$ and the corresponding inferred values of $H_0$. 
EM acknowledges support from the Swedish Research Council under Dnr VR 2020-03384.
AG acknowledges support from the Swedish Research Council under Dnr VR 2020-03444, and the Swedish National Space Board, grant 110-18.
\end{acknowledgements}
\appendix
\begin{figure}[ht]
    \centering
	\includegraphics[width=0.45\linewidth]{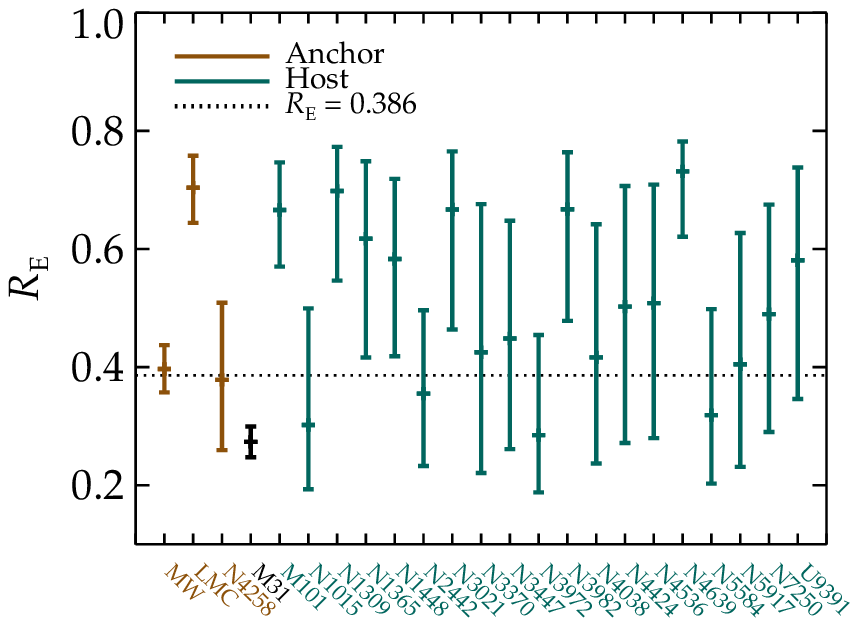}
	\includegraphics[width=0.45\linewidth]{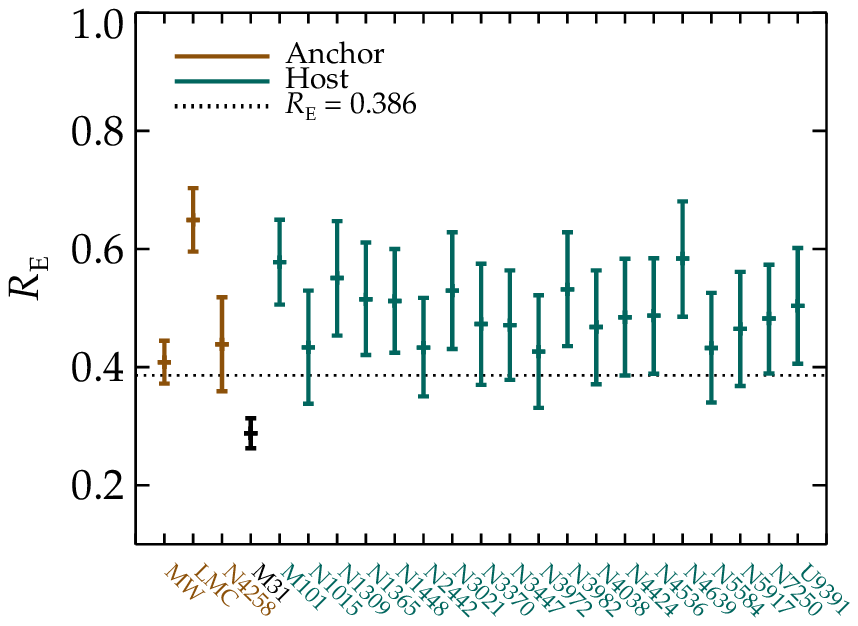}
	\caption{The result of fitting individual galactic values for $R_{\rm E}$ using the color excess Cepheid calibration, imposing flat prior constraints $R_{\rm E}=[0.15,0.8]$ (left panel) and  Gaussian prior constraints $R_{\rm E}=0.48\pm 0.1$ (right panel). Anchor galaxies are denoted in brown and SNIa host galaxies in petrol. The dotted line corresponds to $R_{\rm E} = 0.386$.\label{fig:rhmin015max08}}
\end{figure}

\section{System of Equations}\label{sec:sysofeqns}
Following \citet{Riess:2016jrr}, we collect all data points and their corresponding uncertainties and possible correlations in the matrices ${\bf Y}$ and ${\bf C}$. This includes the Wesenheit magnitudes of all Cepheids, $m^{\rm W}_{{\rm H},i,j}$, including the anchors. The exception is the MW Cepheids for which we use $m_{\pi,j}\equiv m_{{\rm H},j}^{\rm W}-10+\frac{5}{\ln 10}\ln\pi$. Next, we have the measured anchor distances, $\mu_{\rm N4258}, \mu_{\rm LMC}$ and possibly $\mu_{\rm M31}$. Finally, data points include the B-band SNIa magnitudes in the Cepheid hosts, $m_{{\rm B},i}$:
\be
{\bf Y}=\begin{bmatrix} 
m_{\pi,j} \\ m^{\rm W}_{{\rm H},1,j} \\ \vdots \\  m^{\rm W}_{{\rm H},19,j} \\  m^{\rm W}_{{\rm H,N}4258,j}\\  m^{\rm W}_{{\rm H,LMC},j}\\  \mu_{\rm N4258} \\ \mu_{\rm LMC} \\ m_{{\rm B},1} \\  \vdots \\ m_{{\rm B},19} 
\end{bmatrix},
\quad
{\bf C}=\begin{bmatrix} 
\sigma^2(m_{\pi,j}) & 0 & \ldots &&&&&0 \\
0 & \sigma^2(m^{\rm W}_{\rm H}) & 0 & \ldots &&&&\vdots \\
\vdots && \ddots &&&&&\\
&&&\sigma^2(\mu_{\rm N4258}) &&&&\\
&&&&\sigma^2(\mu_{\rm LMC}) &&&\\
&&&&&\sigma^2(m_{{\rm B},1})&&\\
&&&&&&\ddots&\\
0&&&&&&&\sigma^2(m_{{\rm B},19})
\end{bmatrix}
\ee

Collecting the  model parameters in the matrix ${\bf X}$
\be
{\bf X}=\begin{bmatrix}
\mu_{0,1} \\ \vdots \\ \mu_{0,19}\\ \mu_{\rm N4258}\\ \mu_{\rm LMC}\\ M_{\rm H}^{\rm W}\\ b_{\rm W}^{\rm s}\\ b_{\rm W}^{\rm l}\\ Z_{\rm W}\\ zp\\ M_{\rm B} \end{bmatrix},
\ee
we can relate data and parameters through ${\bf Y}={\bf A}{\bf X}$ where in schematic form
\setcounter{MaxMatrixCols}{20}
\be
{\bf A}=\begin{bmatrix} 
0 &\ldots &&&&&& 1 & [{\rm P}]^{\rm s}_{{\rm MW},1} & [{\rm P}]^{\rm l}_{{\rm MW},1} &[{\rm {\rm M}/{\rm H}}]_1  & \frac{-5\pi_1^{-1}}{\ln 10} & 0\\
\vdots &&&&&&&&&&&&\\
0 &&&&&&& 1 & [{\rm P}]^{\rm s}_{{\rm MW},N} & [{\rm P}]^{\rm l}_{{\rm MW},N} &[{\rm {\rm M}/{\rm H}}]_N  & \frac{-5\pi_N^{-1}}{\ln 10} & 0\\
1 &0 &\ldots &&&&& 1 &[{\rm P}]^{\rm s} & [{\rm P}]^{\rm l} & [{\rm {\rm M}/{\rm H}}] & 0 & \vdots\\
\vdots &1 & 0 & \ldots &&&& 1 &\vdots  & \vdots  & \vdots & \vdots & \\
&\ldots &&&& 1 & 0 & 0 &&&&& \\
&&&&&& 1 & 0 &&&&& \\
1 & 0 & \ldots &&&&&&&&&& 1\\
0 & 1 & 0 &\ldots &&&&&&&&& 1\\
\vdots &&&&&&&&&&&&\vdots \\
0 &\ldots&&&1&0 &\ldots&&&&&&1 
\end{bmatrix}.
\ee
Note that the first $N$ rows correspond to equation~\ref{eq:mpi} and so forth, so that ${\bf A}$ is a $n_{\rm param}\times n_{\rm data}$ matrix. We can solve for the parameter matrix ${\bf X}$ and its covariance matrix ${\bf \Sigma}$ analytically 
\be
{\bf \Sigma} = \left[{\bf A}^T{\bf C}^{-1}{\bf A}\right]^{-1},\quad {\bf X}={\bf \Sigma}\left[{\bf A}^T{\bf C}^{-1}{\bf Y}\right].
\ee  
We use the {\tt emcee} Python implementation of the affine-invariant ensemble sampler for Markov chain Monte Carlo (MCMC) \citep{2013PASP..125..306F} to obtain constraints on $R_{\rm W}$ and $R_{\rm E}$. At each step, we recompute the Wesenheit magnitude uncertainties for the specific values of $R_{\rm W}$ and $R_{\rm E}$, and use the formalism described above to solve for the linear parameters in the fit. The exceptions are for fixed values of $R_{\rm W}$, as well as cases where we investigate the bias induced from not correctly taking the color uncertainties into account, in which cases we can extend the linear formalism to include also $R_{\rm W}$ and $R_{\rm E}$. We have checked that in these cases, the linear treatment and the MCMC formalism give consistent results.

\section{Results for Individual Anchors}\label{sec:indanch}
\begin{figure}[ht]
    \centering
	\includegraphics[width=0.32\linewidth]{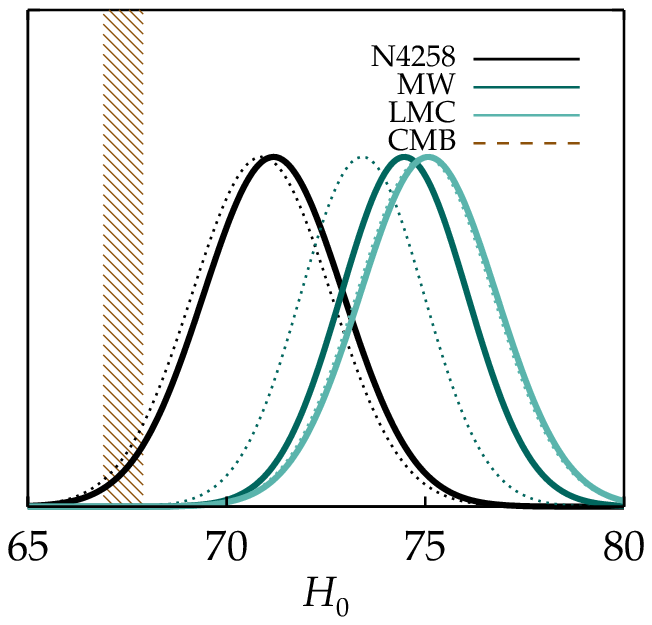}
	\includegraphics[width=0.32\linewidth]{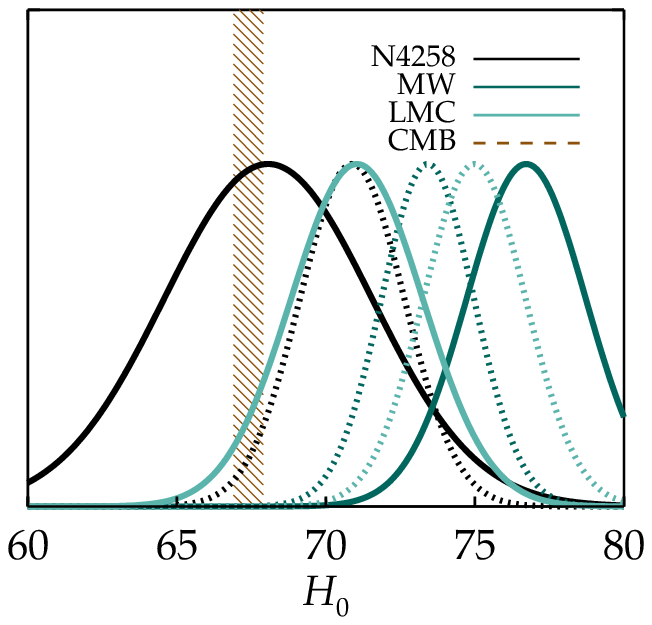}
	\includegraphics[width=0.32\linewidth]{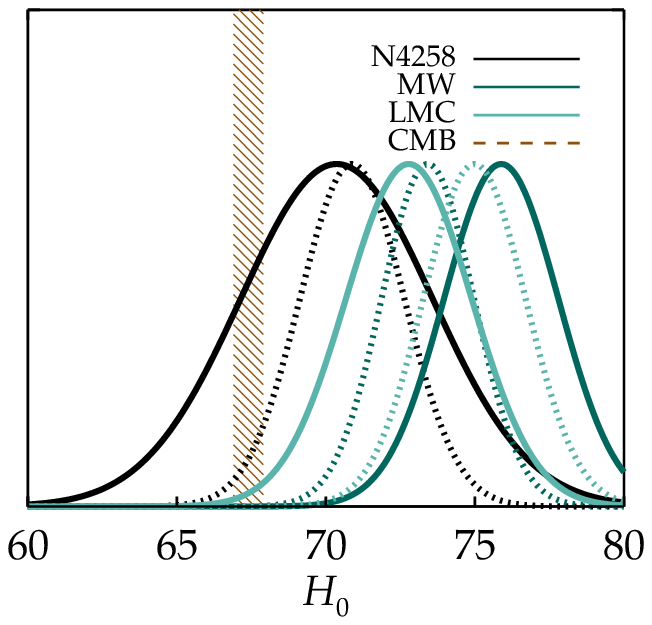}
	\caption{Results for $H_0$ for different anchor distances. Dotted lines are fitted using the Wesenheit calibration with $R_{\rm W}= 0.386$ as in \citet{Riess_2021}. {\em Left panel:} Solid lines are fitted for individual galactic values of $R_{\rm W}$ using Wesenheit magnitudes.  {\em Middle panel:} Solid lines are fitted for individual galactic values of $R_{\rm E}$ using estimated color excesses assuming flat priors in the interval $R_{\rm E}=[0,1]$. Assuming slightly more restrictive priors $R_{\rm E}=[0.15,0.8]$ yield similar results. {\em Right panel:} Solid lines are fitted assuming Gaussian prior values for $R_{\rm E}=0.48\pm 0.1$.\label{fig:anchtension}}
\end{figure}

In Figure~\ref{fig:anchtension}, we show the inferred Hubble constant from each individual anchor. Here, we have used a globally fitted $R_{\rm W}$ for the observed color calibration and individual $R_{\rm E}$ for each galaxy for the color excess calibration. A global P-L relation is assumed in both cases. Dotted lines are for a global value $R_{\rm W}=0.386$ using the Wesenheit calibration, closest resembling the case in \citet{Riess_2021}. The solid lines are for the case of fitting for a global $R_{\rm W}$  and individual $R_{\rm E}$.    

\begin{figure}[ht]
    \centering
	\includegraphics[width=0.8\linewidth]{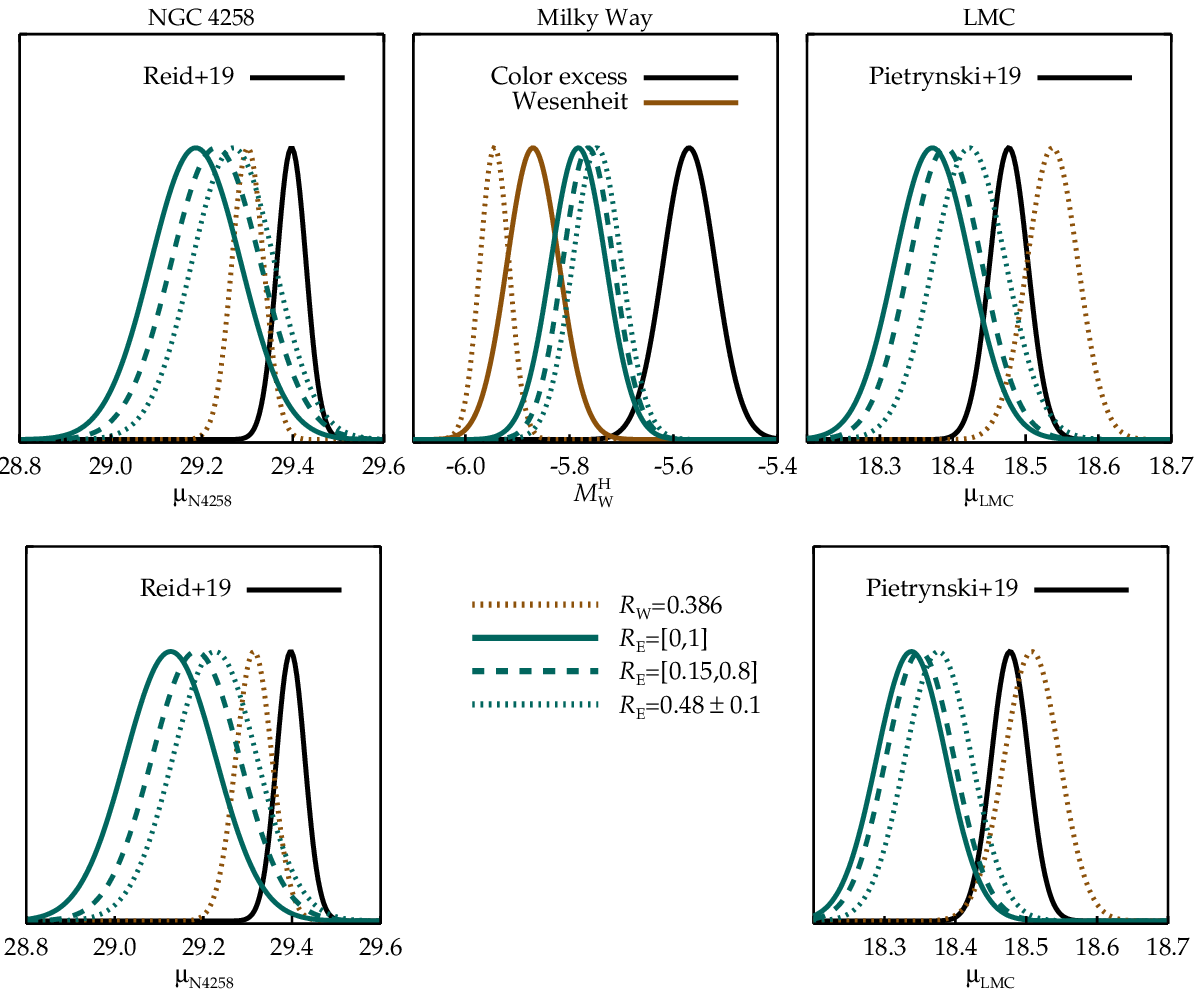}
	\caption{{\em Upper row:} Individual anchor distances derived using the Cepheid and SNIa distance ladder using the other two anchors compared with their
	independently measured distances in \citet{Reid_2019}, \citet{Pietrzy_ski_2019} and using MW Cepheids only to derive $M_{\rm H}^{\rm W}$ (solid black lines, except $M_{\rm H}^{\rm W}$ derived with the Wesenheit calibration using observed colors that is depicted with the solid brown line). {\em Lower row:} The distances to the NGC 4258 and the LMC anchors as inferred using MW data to calibrate the Cepheid magnitudes, compared to their independently measured distances (solid black lines).\label{fig:anchtensionwide}}
\end{figure}

\begin{table}
  \centering
  \caption{The tension between the three different anchors. For the first three columns, the independently measured distance to an anchor is compared to the value derived using the Cepheid and SNIa distance ladder calibrated using the other two anchors. In the case of the MW, the distance is represented by the Cepheid absolute magnitude, $M_{\rm H}^{\rm W}$ derived using MW Cepheids only. In the last column (Milky Way 2), we employ an alternative approach for the Milky anchor where we compare the distances to NGC 4258 and LMC derived using the MW anchor, calibrated using the full Cepheid sample, to their independently measured values. For the anchor with the largest deviation from the other two anchors, we also (in parenthesis) give the significance of deviation corrected for the "look elsewhere" effect.}
  \label{tab:anchtension}
  \begin{tabular}{|c|c|c|c|c|}
    \hline
    Calibration method & NGC 4248  & LMC & Milky Way & Milky Way 2\\
    \hline
$R_{\rm W}=0.386$& $2.0\,\sigma\, (1.6\,\sigma)$ & $1.4\,\sigma$ & $1.3\,\sigma$ & $0.6\,\sigma$\\ 
Global $R_{\rm W}$& $2.1\,\sigma\, (1.7\,\sigma)$ & $1.4\,\sigma$ & $1.1\,\sigma$ & $0.8\,\sigma$\\ 
$R_{\rm E}=[0,1]$& $2.0\,\sigma$ & $1.8\,\sigma$ & $3.0\,\sigma\, (2.6\,\sigma)$ & $3.4\,\sigma\, (3.1\,\sigma)$\\ 
$R_{\rm E}=[0.15,0.8]$& $1.6\,\sigma$ & $1.6\,\sigma$ & $2.7\,\sigma\, (2.4\,\sigma)$ & $3.1\,\sigma\, (2.7\,\sigma)$\\
$R_{\rm E}=0.48\pm 0.1$& $1.3\,\sigma$ & $0.9\,\sigma$ & $2.5\,\sigma\, (2.1\,\sigma)$ & $2.5\,\sigma\, (2.1\,\sigma)$\\
\hline
  \end{tabular}
\end{table}

\section{Model selection}\label{sec:modelselection}
The choice of C-L and P-L calibration method can have a large impact on the inferred $H_0$. Unfortunately, from data alone, it is not obvious which of calibration models is preferred.
As expected, allowing for more freedom in the model, the fit will normally improve, the exception being when fitting for a global $R_{\rm W}$ and $R_{\rm E}$, in which case the Wesenheit magnitude errors are slightly decreased because of the slightly shifted values of $R_{\rm W}$ and $R_{\rm E}$\footnote{When comparing the quality of the fit for different models, we add the same additional scatter of $\sigma_{\rm sc}=0.06$ to the P-L relation. When constraining parameters within a given model, the scatter is adjusted to give a $\chi^2/{\rm dof}=1$.}. In terms of model selection, results are ambiguous, see Table~\ref{tab:IC}. Here, the Akaike Information Criterion (AIC) and the Bayesian Information Criterion (BIC) are defined by
\begin{align}
{\rm AIC}&\equiv 2n_{\rm par}+\chi^2,\nonumber \\
{\rm BIC}&\equiv n_{\rm par}\ln n_{\rm dat}+\chi^2,
\end{align}
with the latter more penalizing for models with extra parameters for $n_{\rm dat}>8$. In terms of the AIC, fitting for individual values of $R_{\rm E}$ with or without individual  $b_{\rm W}$ constitute the preferred models, whereas in terms of BIC, fixed values of $R_{\rm W}=R_{\rm E}=0.386$ together with a global P-L relation yields the best result. 
A more careful analysis, beyond the scope of this paper, would entail calculating the Bayesian evidence by the integrating the prior times the likelihood over
the entire parameter space of the model. In the last column of Table~\ref{tab:IC}, we also account for the $p$-value of the fit.
As it stands, it is unclear to what degree these results provide guidance on the choice of calibration model.

\begin{table}
  \centering
  \caption{Resulting $H_0$ and quality of fit for different color calibration models for Wesenheit calibration, $R_{\rm W}\,({\rm V} - {\rm I})$, and color excess calibration, $R_{\rm E}\,\hat{E}\,({\rm V} - {\rm I})$. $R_{\rm E}=[0,1]$ and $R_{\rm E}=[0.15,0.8]$ correspond to flat prior distributions and $R_{\rm E}=0.48\pm 0.1$ to Gaussian prior distributions.}
  \label{tab:IC}
  \begin{tabular}{|c|c|c|c|c|c|c|}
    \hline
    Calibration method & Added parameters &$H_0$ (Planck tension) &$\chi^2_{\rm min}$     &AIC    &BIC & $p$-value\\
    \hline
    Wesenheit calibration\\
    \hline
$R_{\rm W}=0.386$&	 $-$                   &$73.1\pm 1.3$  $(4.1\,\sigma)$  &1670.5 &1726.5	&1877.8 & $0.17$\\ 
Global $R_{\rm W}$& $R_{\rm W}$                   &$73.2\pm 1.3$  $(4.2\,\sigma)$  &1671.8 &1729.8	&1886.5 & $0.16$\\ 
$R_{\rm W}=0.386$ and individual $b_{\rm W}$& $2\rightarrow 23\; b_{\rm W}$       &$76.5\pm 1.9$  $(4.6\,\sigma)$  &1618.6  &1716.6 &1981.5 & $0.34$\\
Global $R_{\rm W}$ and individual $b_{\rm W}$& $R_{\rm W},\;2\rightarrow 23\; b_{\rm W}$       &$76.5\pm 1.9$  $(4.7\,\sigma)$  &1620.0  &1720.0 &1990.3 & $0.33$\\
    \hline
    Color excess calibration\\
    \hline
$R_{\rm E}=0.386$&	  $-$                       &$73.0\pm 1.3$  $(4.0\,\sigma)$  &1666.1 &1722.1	&1873.5 & $0.22$\\ 
Global $R_{\rm E}$&   $R_{\rm E}$                     &$73.2\pm 1.7$  $(3.3\,\sigma)$  &1667.5 &1725.5	&1882.3 & $0.18$\\ 
$R_{\rm E}=0.386$ and individual $b_{\rm W}$& $2\rightarrow 23\; b_{\rm W}$           &$76.1\pm 1.9$  $(4.4\,\sigma)$  &1615.8	&1713.8 &1978.7 & $0.36$\\ 
Global $R_{\rm E}$ and individual $b_{\rm W}$&  $R_{\rm E},\;2\rightarrow 23\; b_{\rm W}$          &$76.0\pm 2.2$  $(3.9\,\sigma)$  &1617.3	&1717.3 &1987.6 & $0.34$\\ 
Individual $R_{\rm E}=[0,1]$&  $23\; R_{\rm E}$         &$73.9\pm 1.8$  $(3.4\,\sigma)$  &1578.9	&1676.9 &1941.7 & $0.60$\\ 
Individual $R_{\rm E}=[0,1]$ and $b_{\rm W}$& $23\; R_{\rm E},\; 2\rightarrow 23\; b_{\rm W}$ &$75.1\pm 2.3$  $(3.2\,\sigma)$  &1537.8	&1677.8 &2056.2 & $0.73$\\ 
Individual $R_{\rm E}=[0.15,0.8]$&  $23\; R_{\rm E}$         &$73.6\pm 1.8$  $(3.4\,\sigma)$  &1587.5	&1685.5 &1950.4 & $0.54$\\
Individual $R_{\rm E}=0.48\pm 0.1$&  $23\; R_{\rm E}$         &$73.4\pm 1.7$  $(3.4\,\sigma)$  &1622.8	&1720.8 &1985.6 & $0.30$\\
\hline
  \end{tabular}
\end{table}

\bibliography{main}{}
\bibliographystyle{aasjournal}



\end{document}